\documentclass[12pt]{iopart}
\usepackage{graphicx}
\usepackage{amsfonts}
\begin{document}
\title[Spin waves in disordered systems]{Spin waves in disordered magnetic systems }
\author{L.V. Lutsev}
\address{ Research Institute "Ferrite-Domen", Chernigovskaya 8, Saint Petersburg, 196084, Russia}
\ead{lutsev@domen.ru }

\begin{abstract}
Long-wavelength spin waves in disordered magnetic systems have
been investigated. In the framework of the Heisenberg model with
magnetic dipole and exchange interactions between spins it is
found that an additional longitudinal spin wave mode appears. This
mode is characterized by variations of the value of the magnetic
moment density. In order to analyse influence of the magnetic
disorder on spin wave dispersion relations, the special case of
volume and surface spin waves in the Damon-Eshbach (DE) geometry
in films with magnetic disorder is considered. It is revealed
strong influence of the magnetic disorder on surface spin waves,
which consist of two branches -- the DE mode and the longitudinal
spin wave mode. Decrease of the ordering parameters leads to a
decrease of the initial frequency and the curve slope of the DE
mode dispersion curve and to an increase of the dispersion curve
slope of the longitudinal mode. If the spin noncollinearity is
high, then the DE mode dispersion assumes a curve with the
backward character. It is found that the dispersion relation of
the longitudinal mode is weak temperature dependent. The developed
model can explain the observed double-peak structure of FMR
spectra in magnetic nanocomposites.
\end{abstract}

\section{Introduction}

Amorphous magnetic materials and nanocomposites consisted of
magnetic nanoparticles in insulating matrix are important types of
disordered magnetic systems. Due to the amorphous structure,
amorphous magnetic materials possess unusual properties such as
the influence of fluctuations of the exchange interaction on the
magnetic state, the appearance of the roton spin wave branch and
the giant $\Delta E$-effect~\cite{ref1,ref2,ref3,ref4}.
Nanocomposites consisting of ferromagnetic metal nanoparticles
embedded in an insulating matrix exhibit interesting properties
such as giant magnetoresistance due to spin-dependent
tunneling~\cite{ref5,ref6,ref7,ref8,ref9,ref10,ref11}, the
occurrence of additional modes in the ferromagnetic resonance
(FMR) spectrum in a narrow range near the percolation
threshold~\cite{ref12,ref13,ref14}, sharp increase in the FMR
linewidth with decreasing metal content~\cite{ref12,ref13,ref15},
excellent soft magnetic properties governed by the interaction
between ferromagnetic nanoparticles~\cite{ref16,ref17} and the
influence of inter-cluster dipolar interactions on the magnetic
relaxation of nanoclusters~\cite{ref18,ref19}. The effect of the
injection magnetoresistance has been observed on heterostructures
of silicon dioxide with cobalt nanoparticles grown on GaAs
substrates~\cite{ref20,ref21}. Owing to this, these
heterostructures are of great interest for spintronic devices.

In order to analyze the above-mentioned properties and effects, it
is utterly important to get information about magnetic structures
of disordered systems. One of the effective tools, which can be
used for the characterization of magnetic structures is spin waves
propagating in these disordered magnetics. Spin waves are very
sensitive to the magnetic disorder. This exhibits in changes of
spin wave dispersion curves. For investigation of changes of
dispersion curves, one can use spin waves generated by
antennae~\cite{ref22,ref23,ref24} or thermally excited spin waves
observed by the Brillouin light scattering
technique~\cite{ref25,ref26,ref27,ref28,ref29}. In order to
interpret experimental results, it is need to determine
theoretical dependencies between spin wave dispersion curves and
disordered parameters. This leads us to develop a model of spin
waves in disordered systems.

In this paper, in the framework of the Heisenberg model with
magnetic dipole and exchange interactions between spins we
consider long-wavelength spin waves in disordered magnetic
systems. Averaging pseudodifferential Landau-Lifshitz
equations~\cite{ref30}, in section 2 we derive the equation
describing spin waves. In order to analyse influence of the
magnetic disorder on spin wave dispersion relations, in section 3
we consider particular solutions for the case of spin waves in the
Damon-Eshbach (DE) geometry. We have found that in disordered
magnetic systems a longitudinal spin wave mode appears. This mode
is analogous to plasmon modes in solid states~\cite{ref31,ref32}.
In contrast with disordered systems, in ferromagnetically ordered
ones the longitudinal mode is degenerated and is not observed.
Taking into account the existence of the longitudinal spin wave
mode in disordered systems, in section 4 it is supposed that the
double-peak structure of FMR spectra observed in magnetic
nanocomposites~\cite{ref33,ref34,ref35,ref36} can be explained by
excitations of usual spin waves and the longitudinal spin wave
mode.

\section{Derivation of long-wavelength spin wave equation for disordered magnetic systems}

In order to derive equation describing long-wavelength spin waves
in disordered magnetics, we consider the Heisenberg model with
magnetic dipole and exchange interactions between
spins~\cite{ref30}. We will use the approximation of effective
propagators and interaction lines of the spin operator diagram
expansion. In the framework of this approximation the local
dynamics of the spin system is described by Landau-Lifshitz
equations. The ac magnetic field originated from spins is
described by the equation for the magnetostatic potential. The
pseudodifferential Landau-Lifshitz equations have the generalized
form and take into account longitudinal temperature dependent
variations of the variable magnetization. Averaging over a small
volume $\delta V$ and using the transformation from the local
basis to the global one, we will obtain the average value of
variations of the magnetic moment density. Substitution for this
average value in the equation for the magnetostatic potential
gives us the spin wave equation for disordered magnetic systems
for spin waves with wavelength $\lambda\gg(\delta V)^{1/3}$.

\begin{figure*}
\begin{center}
\includegraphics*[scale=.8]{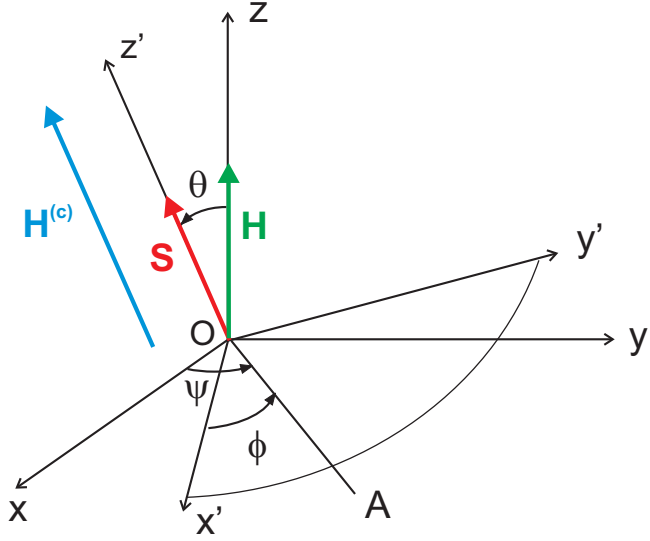}
\end{center}
\caption{Transformation from the local basis $(x',y',z')$ to the
global basis $(x,y,z)$. The axis $OA$ lies in the surface $xOy$.}
\label{Fig1}
\end{figure*}

In order to analyze disordered spin systems and to derive spin
wave equation, we choose the local basis $(x',y',z')$ such that
the axis $Oz'$ is parallel to the self-consistent field $\vec
H^{(c)}=\vec H+\vec H^{(ex)}+\vec H^{(dip)}$ (figure~\ref{Fig1}),
where $\vec H$ is the applied magnetic field, $\vec H^{(ex)}$ is
the exchange field and $\vec H^{(dip)}$ is the dipole magnetic
field originated from neighbouring spins. The exchange field can
be written as $H^{(ex)}_{\mu}(\vec r)=
(g\mu_B)^{-1}\sum_{\vec{r}'} I_{\mu\nu}(\vec
r-\vec{r}')\langle\langle S_{\nu}(\vec{r}')\rangle\rangle$, where
$\mu,\nu =\{{+},{-},z'\}$, $I_{\mu\nu}$ is the exchange
interaction between spins, $g$ is the Land\' e factor, ${\mu}_B$
is the Bohr magneton and $\langle\langle
S_{\nu}(\vec{r}')\rangle\rangle =\{\langle\langle
S_{+}(\vec{r}')\rangle\rangle, \langle\langle
S_{-}(\vec{r}')\rangle\rangle, \langle\langle
S_{z'}(\vec{r}')\rangle\rangle\}$ $(S_{\pm} = S_{x'}\pm iS_{y'})$
is the statistical average spin. It is supposed that the summation
in this relation and in all following relations is performed over
all repeating indices. The dipole magnetic field can be written
as~\cite{ref30}

\begin{equation}
H^{(dip)}_{\mu}(\vec r)=g\mu_B\nabla_{\mu} \sum_{\vec{r}'}
\frac1{|\vec r-\vec{r}'|}\nabla'_{\nu}\langle\langle
S_{\nu}(\vec{r}')\rangle\rangle \label{eq1}
\end{equation}
$$=\frac{g\mu_B}{V_a} \nabla_{\mu} \int_V\frac1{|\vec
r-\vec{r}'|}\nabla'_{\nu}\langle\langle S_{\nu}(\vec{r}')
\rangle\rangle \,d^3r' +H^{(a)}_{\mu}(\vec r),$$

\noindent where the first term is the depolarizing magnetic field
of the continuum magnetic sample, $H^{(a)}_{\mu}(\vec r)$ is the
anisotropy magnetic field, $V_a$ is the atomic volume. Without
action of ac magnetic fields at the site $\vec r$ the average
value of spin $\langle\langle\vec S(\vec r)\rangle\rangle$ is
parallel to the field $\vec H^{(c)}$. In this spin orientation the
local energy minimum is occurred. We assume that in the vicinity
of this local energy point the exchange interaction is isotropic,
$2I^{(0)}_{{-}{+}}= 2I^{(0)}_{{+}{-}}=I^{(0)}_{zz}= I^{(0)}$.
Then, if the ac magnetic field $h_{\nu}(\vec r,\omega)$ acts on
spins, in the local basis $(x',y',z')$ the Landau-Lifshitz
equations give relations between small variations of the magnetic
moment density, $m_{\nu}(\vec r,\omega)= g\mu_B
\delta\langle\langle S_{\nu}(\vec r,\omega)\rangle\rangle/V_a$,
and the field $h_{\nu}(\vec r,\omega)$~\cite{ref30}

\begin{equation}
m_{\pm}(\vec{r},\omega)=2\gamma \hat E_{\pm}^{-1}M(\vec{r})
h_{\mp}(\vec{r},\omega) \label{eq2}
\end{equation}

\begin{equation}
m_z(\vec{r},\omega)= \gamma \hat E_z^{-1} \frac{B^{[1]}(p)}{B(p)}
M(\vec{r}) h_z(\vec{r},\omega), \label{eq3}
\end{equation}

\noindent where $h_{\pm}= 1/2(h_x\mp ih_y)$, $\omega$ is the
frequency, $\delta\langle\langle S_{\nu}(\vec
r,\omega)\rangle\rangle$ is the variation of the statistical
average spin $\langle\langle S_{\nu}(\vec r)\rangle\rangle$,
$\gamma= g\mu_B/\hbar$ is the gyromagnetic ratio and $ M(\vec{r})=
g\mu_B B(p)/V_a$ is the magnetic moment density. Functions

\begin{eqnarray}
B(p)&=& SB_S(Sp)\nonumber\\ B^{[1]}(p)&=& S\frac{\partial
B_S(Sp)}{\partial p} \nonumber
\end{eqnarray}

\noindent are expressed by the Brillouin function $B_S$ for the
spin $S$: $B_S(x)= (1+ 1/2S)\coth [(1+ 1/2S)x] - (1/2S)\coth
(x/2S)$. $p=\beta g\mu_B H^{(c)}_z(\vec r)$, $\beta=1/kT$, $k$ is
the Boltzmann constant, $T$ is the temperature. At low temperature
the derivative of the Brillouin function $B^{[1]}(p)$ tends to 0
and the longitudinal variable magnetization $m_z$ is negligible.
The Landau-Lifshitz operators have the form

$$\hat E_{\pm} m_{\pm}(\vec{r},\omega)= [\gamma
H^{(mag)}(\vec{r})\pm\omega] m_{\pm}(\vec{r},\omega) $$
\begin{equation}+\frac{4\pi\gamma\alpha M(\vec{r})}{V_b}
\sum_{\vec{r'}}\int_{V_b}k^2 \exp[i\vec k(\vec{r}-\vec{r'})]
m_{\pm}(\vec{r'},\omega)\,d^3k \label{eq4}
\end{equation}

\begin{equation}\hat E_z m_z(\vec{r},\omega)= \omega m_z(\vec{r},\omega)
, \label{eq5}
\end{equation}

\noindent where $H^{(mag)}= |\vec H+ \vec H^{(dip)}|$ is the value
of the sum of the external magnetic field and the dipole magnetic
field acting on the spin $\vec S(\vec r)$, $V_b= (2\pi)^3/V_a$,
$\alpha=wV_a/4\pi (g\mu_B)^2$ is the exchange interaction constant
and $w$ is the coefficient in the Fourier transform of the
exchange interaction in the local energy minimum with respect to
the space variables: $\tilde I^{(0)}(\vec k) = \sum_{\vec{r}}
I^{(0)}(\vec r)\exp(-i\vec k\vec{r})= \tilde I^{(0)}(0) -wk^2$.
The Landau-Lifshitz equation (\ref{eq3}) is considered in the
linear approximation with respect to $B^{[1]}(p)/B(p)$, therefore,
we have left the main factor $\omega$ in the operator $\hat E_z$
(\ref{eq5}) and have dropped out factors proportional to
$B^{[1]}(p)$ in this operator. In order to simplify notations, we
write the Landau-Lifshitz equations (\ref{eq2}), (\ref{eq3}) in
the local basis in the short form

$$m_{\nu}(\vec r,\omega)= \hat\chi^{(loc)}_{\nu\mu}h_{\mu} (\vec
r,\omega) ,$$

\noindent where $\hat\chi^{(loc)}_{\nu\mu}$ is the
pseudodifferential tensor operator.

In the considered Heisenberg model the interaction between spins
contains the magnetic dipole interaction. From the form of the
magnetic dipole interaction \cite{ref30} it follows that the ac
magnetic field $h_{\nu}$ originated from neighbouring spins is
magnetostatic, i.e. it is expressed in terms of the magnetostatic
potential $\varphi$: $ h_{\nu}= -\nabla_{\nu}\varphi$. In the
approximation of effective propagators and interaction lines in
the global basis $(x,y,z)$ the ac magnetic field of spin waves is
described by the equation for the magnetostatic potential

\begin{equation}
-\Delta\varphi(\vec{r},\omega)+ 4\pi\nabla_im_i(\vec{r},\omega)
=0, \label{eq6}
\end{equation}

\noindent where the index $i$ is $(x,y,z)$.

Spin wave equation for disordered magnetic systems is derived by
averaging the Landau-Lifshitz equations (\ref{eq2}), (\ref{eq3})
over random variables and by the substitution of the average value
of magnetic moment density variations for $m_i$ in
equation~(\ref{eq6}). In order to substitute $m_i$ in
equation~(\ref{eq6}), we must perform the transformation from the
local basis $(x',y',z')$ to the global basis $(x,y,z)$
(figure~\ref{Fig1}) and average the pseudodifferential tensor
operator $\hat\chi^{(loc)}_{\nu\mu}$ over angles $\theta$, $\psi$,
$\phi$, the exchange interaction $I^{(0)}$ and the value of the
dipole magnetic field $H^{(dip)}=|\vec H^{(dip)}|$. Averaging is
performed in a small volume $\delta V$ of the disordered system.
The volume $\delta V\gg V_a$ and must contain a great number of
spins, but its size is considerably smaller then the size of the
sample. The transformation $U$ between the bases $(x',y',z')$ and
$(x,y,z)$ are expressed through the Euler angles $\theta$, $\psi$
and $\phi$~\cite{ref37}. Without loss of generality, the Euler
angle $\phi$ can be assumed equal to 0. Then, in the global basis
$(x,y,z)$ the average tensor operator $\hat\chi^{(av)}_{ij}$,
which gives relations between $m_i$ and $h_j$, $m_i(\vec
r,\omega)= \hat\chi^{(av)}_{ij}h_j (\vec r,\omega)$, has the form

\begin{equation}
\hat\chi^{(av)}_{ij}= \int\limits_{-\infty}^{\infty}
\int\limits_0^{\infty} \int\limits_0^{\pi}\int\limits_0^{2\pi}
U^{-1}_{i\nu} \hat\chi^{(loc)}_{\nu\mu} U_{\mu j}
u(I^{(0)},H^{(dip)}) f(\theta)
\rho(\psi)\sin\theta\,dI^{(0)}\,dH^{(dip)}\,d\theta\,d\psi ,
\label{eq7}
\end{equation}

\noindent where indices $i$, $j$ are $(x,y,z)$, indices $\nu$,
$\mu$ are $(x',y',z')$ and $u(I^{(0)},H^{(dip)})$, $f(\theta)$,
$\rho(\psi)$ are distributions in the volume $\delta V$.
Distributions must be normalized by relations

$$\int_{-\infty}^{\infty}\int_0^{\infty}
u(I^{(0)},H^{(dip)})\,dI^{(0)}\,dH^{(dip)}= 1$$

$$\int_0^{\pi}f(\theta)\sin\theta\,d\theta= 1$$

$$\int_0^{2\pi}\rho(\psi)\,d\psi= 1.$$

\noindent We suppose that the angle $\psi$ is the random variable
with the distribution $\rho(\psi)= (2\pi)^{-1}$. In this case,
taking into account equations (\ref{eq2}), (\ref{eq3}),
(\ref{eq4}), (\ref{eq5}) and well-known expressions of the
transformation $U$ versus the Euler angles~\cite{ref37}

$$U=\left(\begin{array}{ccc} \cos\psi& \sin\psi&0\\ -\cos\theta
\sin\psi &\cos\theta \cos\psi &\sin\theta\\ \sin\theta
\sin\psi&-\sin\theta\cos\psi&\cos\theta
\end{array}\right) $$

\noindent and integrating over $\psi$ in relation (\ref{eq7}), we
find the average tensor operator

\begin{equation}
\hat\chi^{(av)}= \int\limits_{-\infty}^{\infty}
\int\limits_0^{\infty} \left(\begin{array}{ccc} F\xi+D\eta&
G\zeta&0\\ -G\zeta &F\xi+D\eta &0\\ 0&0& 2F\eta+D(1-2\eta)
\end{array}\right)
u(I^{(0)},H^{(dip)})\,dI^{(0)}\,dH^{(dip)}, \label{eq8}
\end{equation}

\noindent where

$$F=\frac 12\gamma (\hat E^{-1}_{+}+\hat E^{-1}_{-})M(\vec{r})$$

$$G=\frac i2 \gamma (\hat E^{-1}_{+}-\hat E^{-1}_{-})M(\vec{r})$$

$$D=\gamma \hat E_z^{-1} \frac{B^{[1]}(p)}{B(p)} M(\vec{r})$$

$$\xi=\frac 12
\int_0^{\pi}f(\theta)(1+\cos^2\theta)\sin\theta\,d\theta$$

$$\eta=1-\xi$$

$$\zeta=\int_0^{\pi}f(\theta)\cos\theta\sin\theta\,d\theta . $$

\noindent Substituting $m_i=-\hat\chi^{(av)}_{ij}\nabla_j \varphi$
in equation~(\ref{eq6}), we obtain the equation describing spin
waves in disordered magnetic systems

\begin{equation}
\Delta\varphi(\vec{r},\omega)+ 4\pi\nabla_i \hat\chi^{(av)}_{ij}
\nabla_j \varphi(\vec{r},\omega) =0. \label{eq9}
\end{equation}

\noindent In connection with equation~(\ref{eq9}) it is need to
make the following remarks.

(1) Equation~(\ref{eq9}) describes long-wavelength spin waves with
wavelength $\lambda\gg(\delta V)^{1/3}$. Effects related to
short-wavelength spin waves such as the appearance of the roton
spin wave branch and the spin wave localization are not described
by equation~(\ref{eq9}) and are out of our consideration.

(2) Relaxation of spin waves can be taken into account in the
one-loop approximation of the considered Heisenberg model, which
is the next approximation after the approximation of effective
propagators and interaction lines~\cite{ref30}. Self-energy
diagrams containing $n$ loops give correction terms to Green
functions and interaction lines proportional to
$(V_a/R^3_{int})^n$, where $R_{int}$ is the radius of the
interaction between spins. For $V_a/R^3_{int}\ll 1$ one-loop
diagrams give greatest correction terms in comparison with
self-energy diagrams with $n>1$ loops.

It is need to note that in the approximation of effective
propagators and interaction lines the average value of the
magnetic moment density is given as

\begin{equation}
M^{(av)}(\vec r)= \int_{-\infty}^{\infty}\int_0^{\infty} M(\vec r)
\zeta u(I^{(0)},H^{(dip)})\,dI^{(0)}\,dH^{(dip)}, \label{eq10}
\end{equation}

\noindent where in the magnetic moment density $M^{(av)}(\vec r)$
the vector $\vec r$ is the center of the small volume $\delta V$.
At the full randomness of the spin disorder the ordering parameter
$\zeta =0$ and, therefore, $M^{(av)}(\vec r)=0$. But, at the same
time, the tensor operator $\hat\chi^{(av)}$ (\ref{eq8}) takes on
the diagonal form and does not tend to zero.

\section{Spin waves in films with magnetic disorder}

\begin{figure*}
\begin{center}
\includegraphics*[scale=.8]{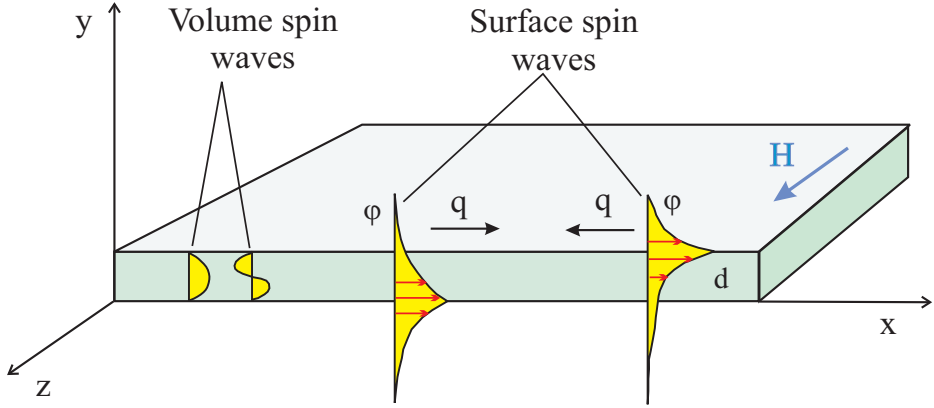}
\end{center}
\caption{Scheme of the film with a magnetic disorder and
distributions of the magnetostatic potential
$\varphi(\vec{r},\omega)$ of volume spin waves and surface spin
waves propagating in opposite directions.} \label{Fig2}
\end{figure*}

In order to clarify peculiarities of long-wavelength spin waves in
disordered magnetic systems, we consider spin waves in the DE
geometry in films with magnetic disorder. In the DE geometry the
orientation of the applied magnetic field $\vec H$ is parallel to
the film surface, spin waves propagate along the axis $Ox$ and the
wavevector $\vec q$ is orthogonal to the field $\vec H$
(figure~\ref{Fig2}). We consider the case, when the spin
wavelength $\lambda\gg(\delta V)^{1/3}$ and the magnetic disorder
is due to a noncollinearity of spins. Therefore, we suppose that
in relations (\ref{eq7}) and (\ref{eq8}) the distribution
$u(I^{(0)},H^{(dip)})$ is equal to $\delta(I^{(0)}-
I^{(0)}_0)\delta(H^{(dip)}- H^{(dip)}_0)$, where $I^{(0)}_0$ is
the average value of the exchange interaction between spins and
$H^{(dip)}_0$ is the average value of the dipole magnetic field
acted from neighbouring spins. Since the depolarizing magnetic
field of the continuum in-plane magnetized film is equal to
zero~\cite{ref23}, in accordance with relation (\ref{eq1}) the
dipole magnetic field $\vec H^{(dip)}_0(\vec r)$ can be reduced to
the anisotropy magnetic field $\vec H^{(a)}_0(\vec r)$. Taking
into account that the film is homogeneous through the thickness
$d$, we will find the solution of equation (\ref{eq9}) in the form

\begin{equation}
\varphi(x,y,z,\omega)=\exp(iqx)\left\{
\begin{array}{ll}
A_1\exp(|q|y),&\qquad y<0\\ A_2\exp(Qy)+A_3\exp(-Qy),&\qquad 0<y<
d\\ A_4\exp(-|q|y),&\qquad y> d
\end{array}\right. \label{eq11}
\end{equation}

\noindent where $Q$ is the transverse wavevector,
$q=2\pi/\lambda$. The magnetostatic potential
$\varphi(\vec{r},\omega)$ and the normal component of the ac
magnetic induction must be continuous at boundaries. This gives
the boundary conditions

$$\left.\varphi(\vec{r},\omega)\right\vert_{+\partial}=
\left.\varphi(\vec{r},\omega)\right\vert_{-\partial} $$

\begin{equation}
\left.4\pi\hat\chi^{(av)}_{yx} \nabla_x \varphi(\vec{r},\omega)+
(1+ 4\pi\hat\chi^{(av)}_{yy}) \nabla_y
\varphi(\vec{r},\omega)\right\vert_{+\partial} \label{eq12}
\end{equation}

$$=\left.4\pi\hat\chi^{(av)}_{yx} \nabla_x
\varphi(\vec{r},\omega)+ (1+ 4\pi\hat\chi^{(av)}_{yy}) \nabla_y
\varphi(\vec{r},\omega)\right\vert_{-\partial},$$

\noindent where $\partial$ is the notation of the boundaries at
$y=0$ and $y=d$. The boundary conditions determine spin wave
dispersion relations. In the DE geometry two types of solutions of
equation (\ref{eq9}) exist -- volume and surface spin waves.

\subsection{Volume spin waves}

Volume spin waves (figure~\ref{Fig2}) are characterized by
imaginary values of the transverse wavevector $Q$. Sewing
$\varphi(\vec{r},\omega)$ in accordance with the boundary
conditions (\ref{eq12}), we determine the coefficients
$A_1,\ldots, A_4$ in the form (\ref{eq11})

$$ A_1=A_4=0 \qquad A_2=-A_3$$

\noindent and get the spin wave dispersion relation

\begin{equation}
\omega_n= \{[\Omega_H+ \alpha\Omega_M(q^2+|Q_n|^2)] [\Omega_H+
\alpha\Omega_M(q^2+|Q_n|^2)+ \xi\Omega_M]\}^{1/2} +\delta\omega_n
, \label{eq13}
\end{equation}

\noindent where the transverse wavevector $Q_n=i\pi n/d$,
$n=1,2,3,\ldots$ is the mode number, $\Omega_H=\gamma H^{(mag)}=
\gamma|\vec H+ \vec H^{(a)}_0|$, $\Omega_M= 4\pi\gamma M$. The
term $\delta\omega_n$ is due to the longitudinal variations of the
magnetic moment density $m_z$ in equation (\ref{eq3}) caused by
nonzero temperature. In the linear approximation with respect to
$B^{[1]}(p)/B(p)$, this term is written as

$$\delta\omega_n =-\frac{\xi\eta B^{[1]}(p)\Omega_M^2}{2B(p)
[\Omega_H+ \alpha\Omega_M(q^2+|Q_n|^2)+ \xi\Omega_M]} .$$

\noindent For the case of ferromagnetic films at zero temperature,
the term $\delta\omega_n =0$, the parameter of ordering $\xi=1$
and relation (\ref{eq13}) coincides with the well-known relation
of spin wave spectra for in-plane magnetized ferromagnetic
films~\cite{ref23}.

\subsection{Surface spin waves}

Surface spin waves (figure~\ref{Fig2}) are characterized by real
values of the wavevector $Q$. Taking into account relation
(\ref{eq8}) and the form of the solution (\ref{eq11}) in the
interior region of the film, from equation (\ref{eq9}) we find
that $Q=q$. Sewing $\varphi(\vec{r},\omega)$ in accordance with
the boundary conditions (\ref{eq12}), we can express the
coefficients $A_1,A_3,A_4$ in the form (\ref{eq11}) versus the
coefficient $A_2$

$$A_3=\frac{(\Omega^2_H-\omega^2)(1+a-\kappa) +\Omega_M(\xi
\Omega_H+ \zeta\omega)}{(\Omega^2_H-\omega^2)(1+a+\kappa)
+\Omega_M(\xi \Omega_H- \zeta\omega)} A_2$$

\begin{equation}
A_1=A_2+A_3 \label{eq14}
\end{equation}

$$A_4=A_2\exp[(|q|+q)d]+ A_3\exp[(|q|-q)d] ,$$

\noindent where the parameter $\kappa=q/|q|=\pm 1$ defines the
direction of the spin wave propagation and

$$a=\frac{\eta B^{[1]}(p)\Omega_M}{B(p)\omega} .$$

\noindent The frequency $\omega$ is determined by the spin wave
dispersion relation

$$\omega= \left\{\Omega_H^2+ \frac{\Omega_M}{8} (4\xi\Omega_H
+\zeta^2 \Omega_Mu)\right.$$

\begin{equation}
\left.\pm \frac{\Omega_M}{8}\left[(4\xi\Omega_H +\zeta^2
\Omega_Mu)^2 +16\Omega_H^2\left(\zeta^2-\xi^2
\right)u\right]^{1/2} \right\}^{1/2} +\delta\omega , \label{eq15}
\end{equation}

\noindent where $u=1-\exp(-2|q|d)$. The term $\delta\omega$ is due
to the longitudinal variations of the magnetic moment density
$m_z$ in equation (\ref{eq3}) and is given by

$$\delta\omega= -\frac{\eta B^{[1]}(p)\Omega_M(\omega^2-
\Omega_H^2)(2\omega^2- 2\Omega_H^2-\xi\Omega_H\Omega_Mu)}{B(p)
\omega^2(8\omega^2- 8\Omega_H^2-4\xi\Omega_H\Omega_M -\zeta^2
\Omega_M^2u)} .$$

\begin{figure*}
\begin{center}
\includegraphics*[scale=.6]{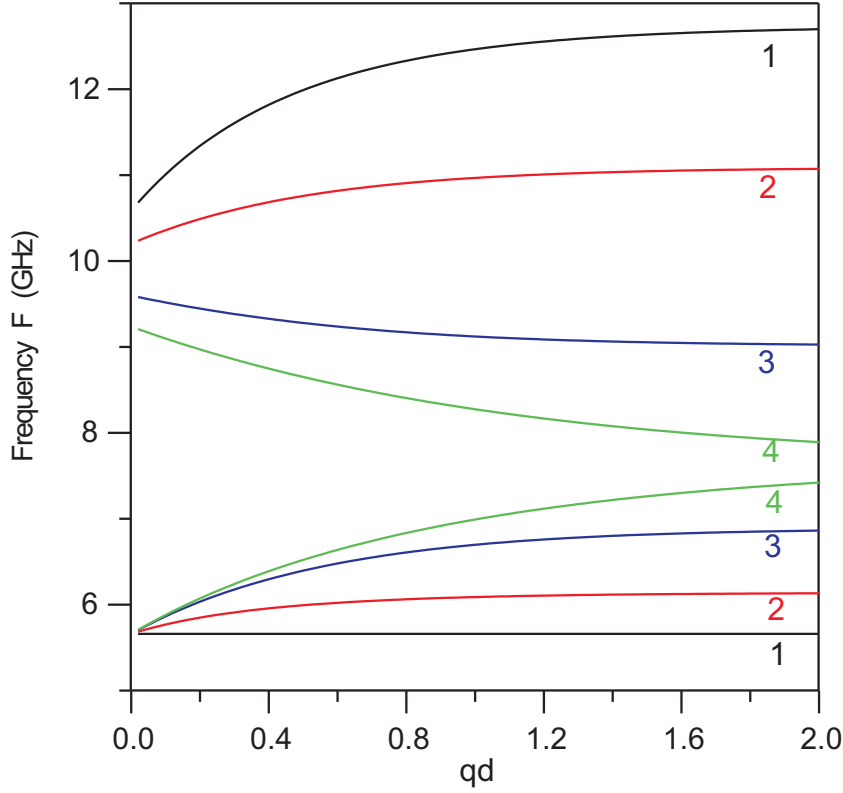}
\end{center}
\caption{Influence of the spin noncollinearity on two branches of
long-wavelength surface spin wave dispersion curves in the film
with a magnetic disorder with the saturation magnetization $4\pi
M= 5$~kOe at the magnetic field $H^{(mag)}=2$~kOe. The upper
branch is the DE mode and the lower branch is the longitudinal
spin wave mode. The wavevector $q$ is normalized by the thickness
$d$. The parameters of ordering are: (1) - $\xi=1$, $\zeta=1$; (2)
- $\xi=0.9$, $\zeta=0.7$; (3) - $\xi=0.75$, $\zeta=0.3$; (4) -
$\xi=2/3$, $\zeta=0$.} \label{Fig3}
\end{figure*}

From relation (\ref{eq14}) one can see that the magnetostatic
potential $\varphi(\vec{r},\omega)$ of surface waves depends on
the direction of the wave propagation -- for the wavevector $q>0$
the wave propagates along one surface and for $q<0$ the wave
propagates along the other surface (figure~\ref{Fig2}). The spin
wave dispersion relation (\ref{eq15}) determines two branches of
surface spin waves. In order to elucidate influence of the spin
noncollinearity on these branches, we have carried out
calculations of dispersion curves (\ref{eq15}) without the term
$\delta\omega$. Dispersion curves are presented in
figure~\ref{Fig3} for the magnetic film with the saturation
magnetization $4\pi M= 5$~kOe for different parameters of
ordering, $\xi$ and $\zeta$. The magnetic field $H^{(mag)}= |\vec
H+ \vec H^{(a)}_0|$, which is the sum of the applied magnetic
field and the anisotropy magnetic field, is equal to 2~kOe for all
curves and the gyromagnetic ratio $\gamma$ is equal to $2\pi\cdot
2.83$ MHz/Oe. Spin waves propagating in opposite directions have
identical dispersion curves. For $\xi=\zeta=1$ spins have
ferromagnetic ordering. In this case, the upper branch is the DE
mode~\cite{ref23}, the lower branch at the frequency $F=\omega
/2\pi= \Omega_H/2\pi= 5.66$ GHz is degenerated. Usually, the lower
branch is not taken into account and is neglected. Decrease of the
ordering parameters $\xi$ and $\zeta$ leads to a decrease of the
initial frequency of the DE mode dispersion curve and to a
decrease of its slope. At the same time, the degeneration of the
lower branch is removed. For films with the spin noncollinearity
of high values, the DE mode dispersion assumes a curve with the
backward character. The limiting case of the spin disorder -- full
randomness -- corresponds to the ordering parameters $\xi=2/3$ and
$\zeta=0$. At these ordering parameters the presented curves in
figure~\ref{Fig3} can be regarded as the limit of surface spin
wave dispersion relations.

\section{Discussion}

\begin{figure*}
\begin{center}
\includegraphics*[scale=.6]{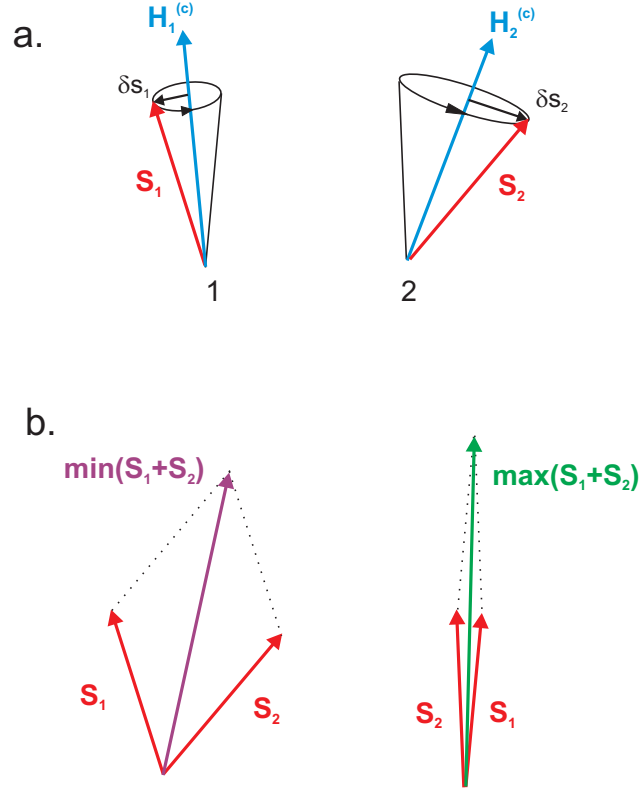}
\end{center}
\caption{(a) Rotation of spins $\vec S_1$ and $\vec S_2$ at
different points 1 and 2 of the space variable in the volume
$\delta V$. (b) Changes in the value of the sum of spins.}
\label{Fig4}
\end{figure*}

Consideration of dispersion curves of long-wavelength spin waves
presented above is carried out for films in the DE geometry and
exhibits two branches of surface spin waves. In consequence of
this consideration, it is arisen the following question: what is
the nature of these two spin wave branches, which are observed in
disordered magnetic systems? It is need to note that only one of
these branches is observed in ferromagnets. In order to explain
this problem, we consider rotation of two noncollinear spins in
the field $\vec H^{(c)}(\vec r)$ in the volume $\delta V$
(figure~\ref{Fig4}a). In the common case, due to different
orientations and values of the field $\vec H^{(c)}(\vec r)$ at
different points 1 and 2 of the space variable $\vec r$,
variations $\delta\vec S_1$, $\delta\vec S_2$ and their phases are
different

$$\vec S_1(t)= \vec S_1^{(0)}+\delta\vec S_1 \exp(i\omega_1t)$$

\begin{equation}
\vec S_2(t)= \vec S_2^{(0)}+\delta\vec S_2 \exp(i\omega_2t) ,
\label{eq16}
\end{equation}

\noindent where $\vec S_1^{(0)}$, $\vec S_2^{(0)}$ are
nonperturbed spins in the points 1 and 2, respectively. Averaging
in relations (\ref{eq7}) and (\ref{eq8}) in the volume $\delta V$
corresponds to the summation of rotating spins. According to
relations (\ref{eq16}), this leads to changes in the value of the
sum of spins during their rotation (figure~\ref{Fig4}b)

\begin{equation}
|\vec S_1(t)+\vec S_2(t)|= [|\vec S_1^{(0)}+\vec S_2^{(0)}|^2
+(\vec S_1^{(0)}\cdot\delta\vec S_2) \exp(i\omega_2t) +(\vec
S_2^{(0)}\cdot\delta\vec S_1) \exp(i\omega_1t)]^{1/2} \label{eq17}
\end{equation}

\noindent and, consequently, the value of the total spin of the
volume $\delta V$ can be changed. For simplicity, in relation
(\ref{eq17}) we restrict ourself by the linear approximation with
respect to $|\delta\vec S_{1(2)}|/|\vec S_{1(2)}|$. From relation
(\ref{eq17}) one can conclude that in disordered systems the
average magnetic moment density $M^{(av)}(\vec r)$ given by
relation (\ref{eq10}) can be variable. This leads to the existence
of spin excitations of two types: (1) rotations of the total spin
of the volume $\delta V$ and (2) variations of the value of this
total spin or variations of the value of the magnetic moment
density $M^{(av)}(\vec r)$. The first excitations are analogous to
excitations in ferromagnets. The second ones can be called as
longitudinal excitations. They are not observed in ferromagnets.
Longitudinal excitations are more expressed in the case of full
spin randomness. In this case, ordering parameters reach the
values: $\xi=2/3$, $\zeta=0$, $\eta=1/3$, and the average tensor
operator $\hat\chi^{(av)}$ (\ref{eq8}) has the diagonal form. Due
to the pole singularity at $\omega= \Omega_H$ in the operator
$\hat\chi^{(av)}$, for the longitudinal spin wave (the lower
branch in figure~\ref{Fig3}) the magnetic moment density
variations $m_i=-\hat\chi^{(av)}_{ij}\nabla_j \varphi$ have high
values.

Longitudinal spin excitations are analogous to plasmon excitations
in solid states. Plasmons are collective oscillations of the
electron density~\cite{ref31,ref32}. In plasma the Coloumb
interaction between electrons is dominant and is determined by the
electrical potential. The longitudinal spin excitations in
disordered magnetic systems are characterized by variations of the
value of the magnetic moment density and are described by the
magnetostatic potential given by equation (\ref{eq9}), which are
analogous to the electron density and to the electrical potential
in plasma, respectively. In this sense, the longitudinal spin
excitations can be called as the plasmon-like spin wave modes.

It is need to note differences between DE and longitudinal modes.
(1) According to spin wave dispersion relation (\ref{eq15}) for
in-plane magnetized films, the frequency $\omega$ of the DE mode
is higher than the frequency of the longitudinal spin wave mode.
This increase of the DE mode frequency is due to the nonzero value
of the saturation magnetization $4\pi M$. This leads to the
nonzero value of $\Omega_M$ in relation (\ref{eq15}). In contrast
with the DE mode, the frequency of the longitudinal mode is close
to $\Omega_H$ determined by the magnetic field $H^{(mag)}$, which
is the sum of the applied magnetic field and the anisotropy
magnetic field. (2) Owing to the inequality $\omega >\Omega_H$ for
the DE mode, the term $\delta\omega$ in the dispersion relation
(\ref{eq15}) depends on temperature through the Brillouin function
factor $B^{(1)}(p)/B(p)$ and the frequency of the DE mode
decreases with temperature increasing. In contrast with this,
since $\omega\to \Omega_H$ for the longitudinal mode, the
dispersion relation of the longitudinal mode slightly depends on
temperature.

FMR spectrum is the limit case of spin waves -- the spin wave
wavevector $q$ is equal to zero. In this limit case, due to the
pole singularity at $\omega= \Omega_H$ in the operator
$\hat\chi^{(av)}$, for the the longitudinal spin wave mode in
disordered magnetic systems the magnetic moment density variations
$m_i=-\hat\chi^{(av)}_{ij}\nabla_j \varphi$ are not zero at
$\nabla_j \varphi\to 0$. This leads to an additional peak in the
FMR spectrum. In magnetic nanocomposites the double-peak structure
of FMR spectra is observed~\cite{ref33,ref34,ref35,ref36}. A weak
secondary peak appears in FMR spectra in nanogranular films
composed of ferromagnetic Fe nanoparticles embedded in SiO${ }_2$
glass matrices~\cite{ref33}. FMR profiles for nanocomposites
Ni/ZnO consisted of nanosized Ni and ZnO particles shows a
secondary peak which appears at the high field tail part of the
absorption curve~\cite{ref35}. Two resonance peaks are observed in
the $\mu^{\prime\prime}$ spectra for Co/ZnO
nanocomposites~\cite{ref34}. Microwave spectra of Ni/ZnO and
Ni/$\gamma$-Fe${ }_2$O${ }_3$ nanocomposites possess a double-peak
behavior in the losses over 2 - 16 GHz frequency
range~\cite{ref36}. Taking into account that the observed effect
can be caused by oxide structures formed in the nanoparticle
interfacial regions, another explanation of the double-peak
structure of FMR spectra can be proposed on the base of the
developed model of spin waves in disordered magnetic systems.
According to this model, one peak is the usual rotation of the
total spin. In the FMR spectra this peak corresponds to usual spin
waves with the zero wavevector, $q=0$. The second peak is due to
variations of the value of the magnetic moment density
$M^{(av)}(\vec r)$. This peak corresponds to the longitudinal spin
wave mode at $q=0$.

\section{Conclusion}

We have investigated long-wavelength spin waves in disordered
magnetic systems. The results of the investigations performed can
be summarized as follows.

(1) In disordered magnetic systems a longitudinal spin wave mode
is appeared. In ferromagnets this mode is degenerated and is not
observed. The longitudinal spin excitation in disordered systems
is characterized by variations of the value of the magnetic moment
density. In this sense, this spin excitation can be called as the
plasmon-like spin wave mode.

(2) The special case of long-wavelength spin waves in the DE
geometry in films with magnetic disorder are considered. It is
found that the magnetic disorder influences volume and surface
spin waves. Surface spin waves have two branches -- the DE mode
and the longitudinal spin wave mode. Decrease of the ordering
parameters leads to a decrease of the initial frequency of the DE
mode, to a decrease of the slope of the DE mode dispersion curve
and to an increase of the dispersion curve slope of the
longitudinal mode. For films with the spin noncollinearity of high
values, the DE mode dispersion assumes a curve with the backward
character. It is found that the dispersion relation of the
longitudinal mode is weak temperature dependent.

(3) The observed double-peak structure of FMR spectra in magnetic
nanocomposites can be explained in the framework of the developed
model.

\section*{Acknowledgment}
This work was supported by the Russian Foundation for Basic Research, grant no 06-02-17030.

\end{document}